\documentclass[twocolumn,aps,epsfig]{revtex4}
\usepackage{graphicx}

\newcommand{\intsum}{\hspace{2mm}\int\hspace*{-5mm}\sum}

\begin{document}

\title{The Casimir Effect in the Presence of a Minimal Length}

\author{U.~Harbach}
\email{harbach@th.physik.uni-frankfurt.de}
\affiliation{Institut f{\"u}r Theoretische Physik\\
J. W. Goethe-Universit\"at\\
Robert-Mayer-Str. 10\\
60054 Frankfurt am Main, Germany}

\author{S.~Hossenfelder}
\email{sabine@physics.arizona.edu}
\affiliation{Department of Physics\\
University of Arizona\\
1118 East 4th Street\\
Tucson, AZ 85721, USA}

\date{\today}

\begin{abstract}
Large extra dimensions lower the Planck scale to values soon
accessible. Not only is the Planck scale the energy scale at which
effects of modified gravity become important. The Planck length
also acts as a minimal length in nature, providing a natural
ultraviolet cutoff and a limit to the possible resolution of
spacetime.

In this paper we examine the influence of the minimal length on the Casimir
energy between two plates.

\end{abstract}

\maketitle

\section{Extra Dimensions}
The study of models with  Large eXtra Dimensions ({\sc LXD}s) has recently received a great deal of attention.
These models,
which are motivated by string theory\cite{Antoniadis:1990ew,Antoniadis:1996hk,Dienes:1998vg}, provide us
with an extension to the standard model (SM) in which
observables can be computed and predictions for tests beyond the
SM can be addressed. This in
turn might help us to extract knowledge about the underlying
theory. The models of {\sc LXD}s successfully fill
the gap between theoretical conclusions and experimental possibilities as the extra hidden
dimensions may have radii large enough to make them
accessible to experiments. The need to look beyond the SM infected many
experimental groups to search for such SM violating processes, for a
summary see e.g. \cite{Azuelos:2002qw}.

Arkani-Hamed, Dimopoulos and Dvali \cite{Arkani-Hamed:1998rs,Antoniadis:1998ig} proposed a solution to the
hierarchy problem by the introduction of $d$
additional compactified spacelike dimensions
in which only the gravitons can propagate. The SM particles
are bound to our 4-dimensional
sub-manifold, often called our 3-brane. Due to its higher dimensional character, the
gravitational force at small distances then is much stronger as it goes in the
radial distance $r$ with the power ${-d-1}$ instead of the usual ${-1}$.
This results in a lowering of
the Planck scale to a new
fundamental scale, $M_{\rm f}$, and gives rise to the exciting possibility of
TeV scale {\sc GUT}s \cite{Dienes:1998vh}. The radius $R$ of the extra dimension
lies in the range mm to $10^3$~fm for $d$ from $2$ to $7$, or the inverse radius
$1/R$ lies in energy range eV to MeV, resp. Throughout this paper  the new
fundamental scale is fixed to be $M_{\rm f}=1$~TeV as a representative value.
For recent constraints see e.g. \cite{Cheung:2004ab}.

\section{The Minimal Length}
Even if a full description of quantum gravity is not yet available, there
are some general features that seem to go hand in hand with all promising candidates
for such a theory. One of them is the need for a higher dimensional spacetime,
one other the existence of a minimal length scale.

As the success of string theory arises
from the fact that interactions are spread out on the world-sheet and do no longer take
place at one singular point, the finite extension of the string has to become important at
small distances or high energies, respectively. Now, that we are discussing the possibility of
a lowered fundamental scale, we want to examine the modifications arising from this as they
might get observable soon. If we do so, we should clearly take into account the minimal length
effects.

In perturbative string theory\cite{Gross:1987ar,Amati:1988tn}, the
feature of a fundamental minimal length scale arises from the fact
that strings cannot probe distances smaller than the string
scale. If the energy of a string reaches this scale
$M_s=\sqrt{\alpha'}$, excitations of the string can occur and
increase its extension\cite{Witten:fz}. In particular, an
examination of the spacetime picture of high-energy string
scattering shows, that the extension of the string grows
proportional to its energy\cite{Gross:1987ar} in every order of
perturbation theory. Due to this, uncertainty in position
measurement can never become arbitrarily small.

Motivations for the occurrence of a minimal length are manifold.
A minimal length can not only be found in string theory but also in loop quantum gravity and
non-commutative geometries. It can be derived from various studies of thought-experiments, from 
the Lie-algebraic stabilisation of the Heisenberg-Poincar{\'e} algebra \cite{Chryssomalakos:2004gk},
from black hole physics, the holographic
principle and further more. Perhaps the most convincing argument, however, is that there seems to
be no self-consistent way to avoid the occurrence of a minimal length scale. For reviews on this topic
see e.g. \cite{Garay:1994en}.

Instead of finding evidence for the minimal scale as has been done in numerous studies, on can
use its existence as a postulate and derive extensions to quantum theories
with the purpose to examine the arising properties in an effective model.

Naturally, the minimum length uncertainty is related to a modification
of the standard commutation relations between position and
momentum \cite{Kempf:1994su}. With the Planck scale as high as $10^{16}$~TeV,
applications of this
are of high interest mainly for quantum fluctuations in the early universe and for
inflation processes and have been examined closely. Now, in the presence of
extra dimensions, we have not only a lowered fundamental scale but also a raised
minimal length.

In \cite{Hossenfelder:2003jz,Hossenfelder:2004up} we used a model for the effects of the minimal
length by modifying the
relation between the wave vector $k$ and the momentum $p$. We assume, no matter how
much we increase the momentum $p$ of a particle, we can never
decrease its wavelength below some minimal length $L_{\mathrm f}$ or, equivalently,
we can never increase
its wave-vector $k$ above $M_{\mathrm f}=1/L_{\rm f}$ \cite{Ahluwalia:2000iw}. Thus, the relation between the
momentum $p$ and the wave vector $k$ is no longer linear $p=k$ but a
function\footnote{Note, that this is similar to introducing an energy
dependence of Planck's constant $\hbar$.} $k=k(p)$.

This function $k(p)$ has to fulfill the following properties:
\begin{enumerate}
\item[a)]  For energies much smaller than the new scale we reproduce the linear relation:
for $p \ll M_{\mathrm f}$ we have $p \approx k$. \label{limitsmallp}
\item[b)] It is an an uneven function (because of parity) and $k \parallel p$.
\item[c)]  The function asymptotically approaches the upper bound $M_{\mathrm f}$. \label{upperbound}
\end{enumerate}
The quantization in this scenario is straight forward and follows the usual procedure.
The commutators between the corresponding operators $\hat{k}$ and $\hat{x}$
remain in the standard form.
Using the well known commutation relations
\begin{eqnarray} \label{CommXK}
[\hat x_i,\hat k_j]={\mathrm i } \delta_{ij}\quad
\end{eqnarray}
and inserting the functional relation between the
wave vector and the momentum then yields the modified commutator for the momentum
\begin{eqnarray} \label{CommXP}
[\,\hat{x}_i,\hat{p}_j]&=& + {\rm i} \frac{\partial p_i}{\partial k_j} \quad.
\end{eqnarray}
This results in the generalized uncertainty relation ({\sc GUP})
\begin{eqnarray} \label{gu}
\Delta p_i \Delta x_j \geq \frac{1}{2}  \Bigg| \left\langle \frac{\partial p_i}{\partial k_j}
\right\rangle \Bigg| \quad,
\end{eqnarray}
which reflects the fact that by construction it is not possible anymore to resolve space-time distances
arbitrarily good. Since $k(p)$ gets asymptotically constant, its derivative $\partial k/ \partial p$
drops to zero and the uncertainty in Eq.(\ref{gu}) increases for high energies.
Thus, the introduction of the minimal length reproduces the limiting
high energy behavior found in string theory \cite{Gross:1987ar}.

The arising physical modifications - as investigated in 
\cite{Hossenfelder:2003jz,Hossenfelder:2004up,Hossenfelder:2004ze} - can be traced back to an effective replacement of the usual momentum measure by a measure which is suppressed at high
momentum
\begin{eqnarray} \label{rescalevolume4}
\frac{{\mathrm d}^{3} k}{(2 \pi)^{3}} \rightarrow \frac{{\mathrm d}^{3} p}{(2 \pi)^{3}}
\Bigg| \frac{\partial k}{\partial p}
\Bigg|  \quad,
\end{eqnarray}
where the absolute value of the partial derivative denotes the Jacobian determinant of $k(p)$.

In the following, we will use the specific relation from \cite{Hossenfelder:2004up} for $k(p)$ by choosing
\begin{eqnarray}
k_{\mu}(p) &=& \hat{e}_{\mu} \int_0^{p} e^{{\displaystyle{-\epsilon p^2}}} \label{model} \quad,
\end{eqnarray}
where $\hat{e}_{\mu}$ is the unit vector in $\mu$ direction, $p^2=\vec{p}\cdot\vec{p}$ and
$\epsilon=L_{\mathrm f}^2 \pi / 4 $ (the factor $\pi/4$ is included to assure, that the limiting value is indeed $1/L_{\mathrm f}$).
Is is easily verified that this expression fulfills the requirements (a) - (c).

The Jacobian determinant of the function $k(p)$ is best computed by adopting spherical coordinates and can be
approximated for $p \sim M_{\mathrm f}$ with
\begin{eqnarray}
\Bigg| \frac{\partial k}{\partial p}
\Bigg|
&\approx& e^{{\displaystyle{-\epsilon p^2}}} \quad.
\end{eqnarray}
With this parametrization of the minimal length effects the modifications read
\begin{eqnarray} \label{gup1}
\Delta p_i \Delta x_i &\geq& \frac{1}{2}  e^{{\displaystyle{+\epsilon p^2}}} \\
\frac{{\mathrm d}^{3} k}{(2 \pi)^{3}} &\rightarrow& \frac{{\mathrm d}^{3} p}{(2 \pi)^{3}}
e^{{\displaystyle{-\epsilon p^2}}}  \quad. \label{gup2}
\end{eqnarray}
In field theory \footnote{For simplicity, we consider a massless scalar
field.}, one imposes the commutation relations Eq.
(\ref{CommXK}) and (\ref{CommXP}) on the field $\phi$ and its
conjugate momentum $\Pi$. Its Fourier expansion leads to the
annihilation and creation operators which must obey
\begin{eqnarray}
\left[\hat{a}_k,\hat{a}^\dag_{k'}\right] &=& - {\rm i}
\left[\hat{\phi}_k,\hat{\Pi}^\dag_{k'}\right]\\
\left[\hat{a}_k,\hat{a}^\dag_{k'}\right] &=& \delta(k-k') \\
\left[\hat{a}_p,\hat{a}^\dag_{p'}\right] &=& e^{{\displaystyle{-\epsilon p^2}}} \delta(p-p')  \quad. \label{coma}
\end{eqnarray}
Note, that it is not necessary for our field to propagate into the extra dimensions to experience
the consequences  of the minimal length scale. In particular, we will assume that the field is bound
on our submanifold to exclude the additional presence of KK-excitations. The existence of the extra dimensions
is important for the case under discussion only by lowering the Planck scale and raising the minimal length.

\section{The Casimir Effect}

Zero-point fluctuations of any quantum field give rise to
observable Casimir forces if boundaries are present
\cite{Casimir:dh}. The Casimir effect is our experimental grip to
the elusive manifestations of vacuum energy. Its importance for
the understanding of the fundamental laws of quantum field theory
lies in the direct connection to the problem of renormalization.
Vacuum energies in quantum field theories are divergent. The
presence of infinities in physics always signals that we have
missed some crucial point in our mathematical treatment.

The Casimir effect has received great attention also in the
context of extra dimensions and has been extensively studied in a
wide variety of topics in those and related scenarios:
\begin{itemize}
\item The question how vacuum fluctuations affect the stability of extra dimensions has been
explored in \cite{Ponton:2001hq,Hofmann:2000cj,Huang:2000qc,Brevik:2000vt,Graham:2002xq,Saharian:2002bw,Elizalde:2002dd,Saharian:2003qp,Setare:2004ni}.
Especially the detailed studies  in the Randall-Sundrum model have shown the major contribution of
the Casimir effect to stabilize the radion \cite{Garriga:2002vf,Pujolas:2001um,Flachi:2001pq,Goldberger:2000dv}.
\item Cosmological aspects like the cosmological constant as a manifestation of the Casimir energy
or effects of Casimir energy during the  primordial cosmic inflation have been analyzed
\cite{Nojiri:2000bz,Peloso:2003nv,Elizalde:2000jv,Pietroni:2002ey,Setare:2003ds,Melissinos:2001fm,Gardner:2001fz,Milton:2001np,Carugno:1995wn,Naylor:2002xk}.
\item The Casimir effect in the context of string theory has been investigated
in \cite{Fabinger:2000jd,Gies:2003cv,Brevik:2000fs,Hadasz:1999tr}.
\item The Casimir effect in a model with minimal length based on the assumption of Path Integral Duality \cite{Padmanabhan:1996ap,Padmanabhan:1998yy}
has been studied in \cite{Srinivasan:1997rs}.
\item It has been shown \cite{Nugaev:1979fj,Nugaev:1984mz} that the Casimir effect provides an analogy to the 
Hawking radiation of a black hole. The presence of Large eXtra Dimensions allows
black hole creation in colliders\cite{Kanti:2004nr} and the understandig of the evaporation properties
is crucial for the interpretation of the signatures.
\end{itemize}

As one might expect, the introduction of a minimal length scale yields an ultraviolet cut off for the
quantum theory which
renders the occurring infinities finite.

Using the above framework, in the presence of a minimal length the operator for the field
energy density is now given by
\begin{eqnarray}
\hat{H}=\frac{1}{2} \intsum {\mathrm d}^{3} p \; \left( \hat{a}^{\dag}_p \hat{a}_p
+ \hat{a}_p \hat{a}^{\dag}_p \right) E \quad,
\end{eqnarray}
where $E$ is the energy of a mode with momentum $p$. With Eq. (\ref{coma}) and $\hat{a}^{\dag}_p \vert 0 \rangle =0$ this results in the
expectation value for the vacuum energy density
\begin{eqnarray} \label{H}
\langle 0 \vert \hat{H} \vert 0 \rangle = \frac{1}{2} \intsum {\mathrm d}^{3} p \; e^{{\displaystyle{-\epsilon p^2}}} E \quad.
\end{eqnarray}
For Minkowski space in $3+1$ dimensions without boundaries, this energy density now is finite due to the
squeezed momentum space at high energies and given by
\begin{eqnarray} \label{mink}
\epsilon_{\mathrm{Mink}} = \langle 0 \vert \hat{H} \vert 0 \rangle = \frac{16}{\pi} \frac{M_{\rm f}}{L_{\rm f}^3} \quad.
\end{eqnarray}
We will now consider the case of two conducting parallel plates in a distance $a$ in direction $z$. We will
neglect effects arising from surface corrections and finite plate width. We will further 
assume that the plates are perfect conductors and infinitely extended in the 
longitudinal directions $x$ and $y$,  such that no boundaries effects are present.
 
The quantization of the wavelengths between the plates in the $z$-direction
yields the condition $k_l = l/a$. Since the wavelengths can no longer get arbitrarily
small, the smallest wavelength possible belongs to a finite number of nodes
$l_{\rm max}=\lfloor a / L_{\rm f} \rfloor$, where the brackets denote the next smaller
integer. Resulting from this, momenta come
in steps $p_l = p (k_l)$ which are no longer equidistant $\Delta
p_l = p_l - p_{l-1}$. Then
\begin{eqnarray} \label{Hplates}
\epsilon_{\mathrm{Plates}}=\pi \hspace*{-2mm} \sum_{l=-l_{\rm max}}^{l_{\rm max}} \hspace*{-2mm}
\Delta p_l \int_0^{\infty} {\mathrm d} p_{\parallel}   \; \;
e^{{\displaystyle{-\epsilon p_{\parallel}^2}}}
e^{{\displaystyle{-\epsilon p_l^2}}}  E~p_{\parallel} \quad,
\end{eqnarray}
where $p_{\parallel}^2=p_x^2 + p_y^2$ and $E^2 = p_{\parallel}^2 +p_l^2$.

Experiments do not measure absolute energy values but 
only differences. Therefore, the difference between the 
inside and the outside region has to be taken, i.e. Eq. (\ref{mink}) has to be subtracted from 
Eq.(\ref{Hplates}). This then yields the Casimir energy accessibly by experiment through
the induced pressure which results in a force acting on the plates. For the
case of two parallel plates, the pressure is negative in the inside, or the force is
attractive, respectively.

In the limit of large $M_{\rm f}$, i.e. of small $L_{\rm f}$, the renormalized
standard result is obtained. This can be seen directly from taking the
difference between the outside and inside region, that is Eq. (\ref{Hplates}) and Eq. (\ref{H}), and
applying the Abel-Plana-formula \cite{Saharian:2000xx}.
In this expression, the integral over the directions parallel to the plates is the same in both
terms and may thus be taken conjoined:

\begin{eqnarray}
\lim_{L_{\rm f}\rightarrow 0}\int_0^\infty \hspace*{-2mm}{\mathrm d} p_\parallel \bigg ( \sum_{l=-l_{\rm max}}^{l_{\rm max}} \Delta p_l~
 e^{\displaystyle{-\epsilon p_l^2}} E~ p_\parallel
 &-& \nonumber \\ 
\int_{\infty}^{\infty} && \hspace*{-8mm}
{\mathrm d} p ~ e^{\displaystyle{-\epsilon p^2}} E~ p_\parallel \bigg ) e^{{\displaystyle{-\epsilon p_{\parallel}^2}}}\nonumber \\
 =  
\lim_{L_{\rm f}\rightarrow 0}\int_0^\infty \hspace*{-2mm}{\mathrm d} p_\parallel \bigg ( \sum_{l=-\infty}^{\infty} \Delta p_l~ 
e^{\displaystyle{-\epsilon p_l^2}} E~p_\parallel 
&-& \hspace*{-1mm} \int_{-\infty}^{\infty} \hspace*{-2mm} {\mathrm d} p~
e^{\displaystyle{-\epsilon p^2}} E~ p_\parallel  \nonumber \\
- 2 \sum_{l=l_{\rm max}}^{\infty}   \Delta p_l~  
e^{\displaystyle{-\epsilon p_l^2}} E~p_{\parallel} \bigg )
e^{{\displaystyle{-\epsilon p_{\parallel}^2}}}\hspace*{-8mm} &&\quad.
\end{eqnarray}

\begin{figure}
\vspace*{-0mm} \centering
\includegraphics[width=3.6in]{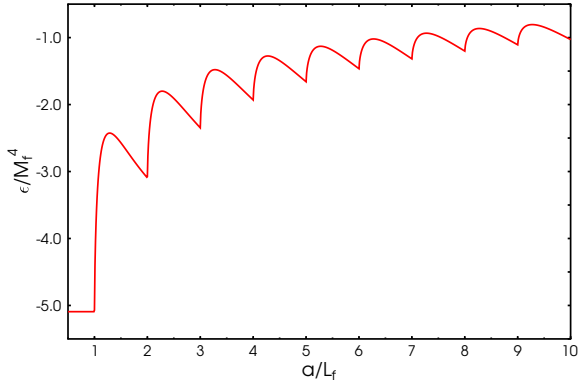} \caption{The Casimir
energy density between two plates of distance $a$ in units of the
minimal length. \label{fig1}}
\end{figure}

Taking the limit $L_{\rm f} \to \infty$ we have $\Delta p_l \to 1/a$ and $l_{\rm{max}} \to \infty$. 
Then, the last
term vanishes, while the first
terms are the same that appear in the classical calculation of the
Casimir energy. Since the exponential, which acts as a dampening
function, is holomorphic \footnote{We take $p^2$ to be $p\cdot p$,
not $p^\ast\cdot p$}, the Abel-Plana-formula can be used to
evaluate the difference. The obtained
integral is uniformly convergent, and one can perform the limit
before the integration. This then yields the classical expression:
\begin{eqnarray}  
\frac{1}{a}\int_0^\infty {\mathrm d} p_\parallel \sum_{l=-\infty}^{\infty} E~p_\parallel
-\int_0^\infty {\mathrm d} p_\parallel \int_{-\infty}^{\infty} {\mathrm d}p~E~p_\parallel\quad. 
\end{eqnarray}
These computations shows very nicely, how the minimal length acts as a natural regulator in
calculating the Casimir energy.

The result of our computation from a numerical analysis is 
shown in Fig. \ref{fig1}. As can be seen, the slope of the curve changes every
time another mode fits between the plates. Although the slope (and thus the Casimir
force) is singular at these points, the plot clearly shows that a finite energy is
sufficient to surmount them and thus the result is physical. Also, the singularities
seem to stem from the assumption of two strictly localised plates and should be cured in
a full theory by the minimal length uncertainty on their position.

If the distance eventually drops below the minimal length, the energy density, and thus the pressure acting
on the plates, becomes constant. This is to be contrasted with the standard
result in which the curve diverges towards minus infinity for small distances.

Though the here discussed minimal length is some orders of magnitude out
of range for experimentally measuring the modifications of the Casimir pressure,
this result is interesting not only from a theoretical point of view: As mentioned
before, the analogy to the black hole's temperature is an important application.
We can state that towards small black hole sizes the temperature 
does not increase according to the Hawking evaporation but is severely modified
close to the new fundamental scale and eventally gets constant. Since the
time evolution of the temperature is mostly ignored for the event generation of 
black hole decays (see e.g. \cite{Tanaka:2004xb}), the here presented result justifies 
this treatment.
\vspace*{.5cm}

\section{Conclusion}

We have discussed the existence of a minimal length scale and used an effective 
model to include it into todays quantum theory. 
Such a minimal scale would affect experimental measurements in the presence 
of Large eXtra Dimensions and yield to interesting phenomenological implications. 
The introduced minimal length acts as a natural ultraviolet regulator of the theory. 
We applied our model to the calculation of the Casimir energy and gave a 
numerical evaluation of the resulting expression. Furthermore, we showed how 
the minimal scale provides a physical motivation for the dampening function method 
used in the classical calculation of the Casimir energy via the Abel-Plana formula.
Using the analogy to the black hole evaporation characteristics we showed that
the time evolution of the system can be ignored close to the new fundamental scale.
 
\section*{Acknowledgments}

This work was supported by the German Academic Exchange Service
({\sc DAAD}), NSF PHY/0301998, {\sc DFG} and the Frankfurt Institute for Advanced Studies ({\sc FIAS}).
We want to thank Marcus Bleicher, Achim Kempf and J{\"o}rg Ruppert for helpful discussions.

\end{document}